\documentclass[]{article}
\usepackage[margin=2.5cm]{geometry}

\usepackage{xr-hyper} 
\usepackage[toc,page]{appendix}
\usepackage{hyperref} 
\usepackage{amsmath,amsfonts,breqn,comment,cancel,mathtools,amssymb,extarrows,mathtools,graphicx,subfigure,setspace,bbold,physics}
\usepackage{amssymb}
\usepackage{url}
\usepackage{cite}
\usepackage{tikz}

\begin{document}

\begin{flushright}
IPM/P-2024/17\\
\end{flushright}
\vspace{.4cm}

\begin{center}{\Large \textbf{
Entanglement in Lifshitz Fermion Theories
}}\end{center}

\vspace*{15mm}
		\vspace*{1mm}

\begin{center}
{Mohammad Javad Vasli$^{a,b}$,  Komeil Babaei Velni$^{a}$, 
  M. Reza Mohammadi Mozaffar$^a$
and  Ali Mollabashi$^b$ 
}
\end{center}
\vspace*{1cm}

\begin{center}
{\it $^a$ Department of Physics, University of Guilan, P.O. Box 41335-1914, Rasht, Iran\\
  $^b$ School of Physics, Institute for Research in Fundamental Sciences (IPM), P.O.Box 19395-5531, Tehran, Iran
		}
		
		\vspace*{0.5cm}
		{E-mails: {\tt vasli@phd.guilan.ac.ir,\{babaeivelni, mmohammadi\}@guilan.ac.ir, mollabashi@ipm.ir}}%
\end{center}


\vspace*{1cm}
\begin{abstract}
We study the static entanglement structure in (1+1)-dimensional free Dirac-fermion theory with Lifshitz symmetry and arbitrary integer dynamical critical exponent. This model is different from the one introduced in [Hartmann et al., SciPost Phys. 11, no.2, 031 (2021)] due to a proper treatment of the square Laplace operator. Dirac fermion Lifshitz theory is local as opposed to its scalar counterpart which strongly affects its entanglement structure. We show that there is quantum entanglement across arbitrary subregions in various pure (including the vacuum) and mixed states of this theory for arbitrary integer values of the dynamical critical exponent. Our numerical investigations show that quantum entanglement in this theory is tightly bounded from above. Such a bound and other physical properties of quantum entanglement are carefully explained from the correlation structure in these theories. A generalization to (2+1)-dimensions where the entanglement structure is seriously different is addressed. 
\end{abstract}


\newpage
\vspace{10pt}
\noindent\rule{\textwidth}{1pt}
\tableofcontents
\noindent\rule{\textwidth}{1pt}
\vspace{10pt}

\section{Introduction}
\label{sec:intro}
With regard to the question about bipartite entanglement structure in quantum field theories and many-body systems, we study unknown features of bipartite entanglement in free fermionic Lifshitz theories. Lifshitz theories are non-relativistic theories that are invariant under
\begin{equation}\label{eq:LifScaling}
t\to\lambda^z t\;\;\;\;\;\;\;,\;\;\;\;\;\;\;\vec{x}\to\lambda\vec{x}\;,
\end{equation}  
where $\lambda>0$ and $z\neq 1$ ($z=1$ corresponds to relativistic theories).

Lifshitz symmetry algebra encompasses generators for the aforementioned scaling symmetry together with rotations and translations (see e.g. \cite{Hartnoll:2009sz}). Compared to relativistic scale invariance, there is no counterpart for the boost generator in Lifshitz algebra. Hence, in contrast with the relativistic case in (1+1)-dimensions (2$d$) (see e.g. \cite{Calabrese:2009qy}), the symmetry is not generally powerful enough to lead to universal entanglement structures. However people have utilized various methods to study the bipartite entanglement structure in such theories in (2+1)-dimensions (3$d$) \cite{Ardonne:2003wa,  Fradkin:2006mb, Hsu:2008af, Fradkin:2009dus, Oshikawa:2010kv, Zaletel:2011ir, Zhou:2016ykv, Chen:2016kjp, MohammadiMozaffar:2017nri, MohammadiMozaffar:2017chk, Angel-Ramelli:2020wfo, Angel-Ramelli:2020xvd}, in 2$d$ \cite{Chen:2017txi, MohammadiMozaffar:2017nri, He:2017wla, MohammadiMozaffar:2017chk, MohammadiMozaffar:2018vmk, Angel-Ramelli:2020wfo, Hartmann:2021vrt, Mozaffar:2021nex}, as well as generalizations to higher dimensions \cite{Angel-Ramelli:2019nji}.\footnote{See also \cite{Solodukhin:2009sk, Rajabpour:2014osa, Mintchev:2022yuo, Berthiere:2023bwn, Boudreault:2021pgj} for related works.}
	
Specifically, entanglement entropy of the ground state of a proposed free Dirac-Lifshitz fermion theory has been partially studied in \cite{Hartmann:2021vrt}. The proposed theory in \cite{Hartmann:2021vrt} leads to a product state as the vacuum state of Lifshitz fermions for \textit{even} values of the dynamical critical exponent\footnote{For a similar symmetry-based argument in Galilean field theories see \cite{Hason:2017flq}.}, while for odd values of $z$ they find $z$-\textit{independent} entanglement entropy inherited from the $z$-independence of the two-point correlation functions.

In this paper, we carefully consider a revisited version for free Dirac-Lifshitz fermion theory (see Eq. \eqref{LII})\footnote{This theory has been previously introduced in \cite{Mozaffar:2021nex}, though the lattice regularization in this paper is different.} which includes a negative momentum mode corresponding to all positive momentum modes for any integer value of the dynamical critical exponent. Indeed, for odd values of $z$, this revisited theory coincides with that in \cite{Hartmann:2021vrt}. On the other hand, for even values of $z$, these theories are drastically different in the sense that the Fermi surface is empty in the one introduced in \cite{Hartmann:2021vrt} but is populated in the revisited version. As a result, we find non-trivial two-point functions and hence quantum and classical correlations in the vacuum state. We study the entanglement structure in certain pure state (basically the vacuum state) and mixed states in this family of theories.   


The rest of the paper is organized as follows: In the next subsection we review our calculation method for entanglement measures stressing on the less well-known part of the numerical method corresponding to logarithmic negativity for fermionic Gaussian states. In section \ref{liffer1+1} we introduce the revisited free Dirac-Lifshitz fermion theory in 2$d$ and discuss the structure of equal-time two-point functions. In section \ref{sec:Numerics} we present our numerical results for entanglement entropy, logarithmic negativity, and address generalizations to higher dimensions which seriously changes the entanglement structure. We also discuss about entropic $c$-theorem counterparts for these non-relativistic theories. The last section is devoted to concluding remarks. 

\subsection*{Technical Preliminaries}
Through this paper we will compute different entanglement and correlation measures employing the correlator method, sometimes referred to as the fermionic covariance matrix formalism \cite{Peschel:2002yqj, Vidal:2002rm, Latorre:2003kg, Casini:2009sr, Eisler:2009vye}. The first measure is entanglement entropy associated to a subregion $A$ which is defined as the von Neumann entropy $S_A=-\mathrm{Tr}\left(\rho_A\log\rho_A\right)$ where $\rho_A$ is the reduced density matrix of subregion $A$ obtained by tracing out the degrees of freedom of the complement of this region. Within the correlator method formalism this quantity can be computed as
	\begin{eqnarray}\label{EE}
		S_A=-\sum_{n=0}^{n_A}(1-\nu_n)\log(1-\nu_n)+\nu_n \log \nu_n,
	\end{eqnarray}	
where $n_A$ is the number of lattice sites included in region $A$	and $\nu_n$'s
	are eigenvalues of the fermionic two point function  
	\begin{eqnarray}
		C_{rs}=\langle  \Psi_r \Psi^{\dagger}_s \rangle,
	\end{eqnarray}
and $\Psi_r$ denotes the fermionic field at site $r$. Moreover, all quantities which are defined as combinations of von Neumann entropies such as mutual information between two subregions $A$ and $B$ defined as
	\begin{eqnarray}\label{MI}
		I(A:B)=S_A+S_B-S_{A\cup B},
	\end{eqnarray}	
can be computed using Eq. \eqref{EE}.

Although the von Neumann entropy is a unique measure for quantum correlations in pure states, the story is complicated for mixed states. It has been shown by \cite{Peres:1996dw, Horodecki:1996nc} that the negative eigenvalues of the partial transposed density matrix measure quantum correlations for mixed states. Following this observation negativities were introduced as computable measures for quantum correlations for mixed states \cite{Vidal:2002zz}. Later it has been shown that this measure is not ideal but a good measure \cite{Plenio:2005cwa, Plenio:2007zz} though widely studied in the literature.

We are interested in logarithmic negativity that is defined as $\mathcal{E}(A:B)=\log||\rho^{T_A}_{AB}||$ where $\rho_{AB}^{T_A}$ is the partial transposed density matrix with respect to $A$ and $||\bullet||$ denotes the trace norm. In conformal field theories that the case of $z=1$ would be an example of, the replica trick and conformal symmetry has been used to compute this quantity for certain configurations, namely a single interval and two adjacent intervals in the vacuum state in \cite{Calabrese:2012nk, Calabrese:2012ew} (for extension to massive case see \cite{Blondeau-Fournier:2015yoa}) and a single interval in finite temperature state in \cite{Calabrese:2014yza}.

In the following, our focus is on Gaussian states in free fermion theories. Let us mention that for bosonic Gaussian states the partial transposition has a simple realization in terms of the covariance matrix \cite{Simon:1999lfr} which leads to straightforward calculation of logarithmic negativity in lattice bosonic models (see e.g. \cite{Calabrese:2012nk, Audenaert:2002xfl, MohammadiMozaffar:2017chk}). In contrast, for fermionic Gaussian states a similar calculation has been a long standing challenge since the partial transposition of a Gaussian density matrix does not preserve the Gaussianity of the state \cite{Eisler:2015tgq}. There has been several attempts to get around this problem \cite{Eisler:2015tgq, Coser:2015mta, Coser:2015eba, Coser:2015dvp, Herzog:2016ohd}\footnote{Prior to these developments, Monte-Carlo and tensor network methods have been utilized to calculate (logarithmic) negativity in one dimensional Ising model \cite{Wichterich:2008vfx, Alba:2013mg, Calabrese:2013mi}.}.  

The authors of \cite{Herzog:2016ohd} suggested a path integral \textit{estimation} to calculate negativities for free fermions. Later on the authors of \cite{Shapourian:2018ozl, Shapourian:2016cqu} introduced a method which can be thought of as a sharp justification for the estimation of \cite{Herzog:2016ohd}, based on concerns about the separability criteria for fermionic states. It has been shown in \cite{Shapourian:2018ozl} that taking the fermion number conservation into account leads to a slightly different separability criteria for fermionic states.\footnote{We thank Ken Shiozaki and Hassan Shapourian for fruitful discussions on this point.} From the separability criteria point of view, this leads to a different partial transposition operation which is denoted by partial time-reversal transpose or \textit{fermionic partial transpose} (for details see \cite{Shapourian:2018ozl,Shapourian:2018lsz}). Here we use this definition to calculate logarithmic negativity for fermionic Gaussian states in our quadratic fermionic theories.

Suppose $A=A_1\cup A_2$ and we are considering the fermion partial transpose over $A_1(A_2)$. The recipe for this calculation is formulated in terms of the covariance matrix given by
\begin{eqnarray}
\Gamma=\mathbf{1}-2C=\Gamma=\begin{pmatrix}
			\Gamma^{11} &\Gamma^{12}\\
			\Gamma^{21} &\Gamma^{22}\\
		\end{pmatrix}.
\end{eqnarray} 
The fermionic partial transposed correlation matrix $C_{\Xi}$ can be evaluated as 
	\begin{eqnarray}
		C_{\Xi}=\frac{1}{2} (\mathbf{1}-(\mathbf{1}+\Gamma_+\Gamma_-)^{-1}(\Gamma_++\Gamma_{-})),
	\end{eqnarray}
where $\Gamma_+$ and $\Gamma_-$ are defined by
	\begin{eqnarray}
		\Gamma_{\pm}=\begin{pmatrix}
			-\Gamma^{11} &\pm i\Gamma^{12}\\
			\pm i \Gamma^{21} &\Gamma^{22}\\
		\end{pmatrix}.
	\end{eqnarray}
Finally, the logarithmic negativity can be computed as 
	\begin{eqnarray}\label{LN}
		\mathcal{E}(A_1:A_2)=\sum_{n}\ln (\xi^{\frac12}_j+(1-\xi_j)^{\frac{1}{2}})+\frac12 \sum_j \ln (\zeta_j^2+(1-\zeta_j)^2),
	\end{eqnarray}
	where $\zeta_j$ and $\xi_j$ are eigenvalues of $C$ and $C_{\Xi}$ respectively.
	
\section{Fermionic Lifshitz Theory}\label{liffer1+1}

In this section we will introduce a revisited free Dirac-Lifshitz fermion theory in 2$d$ and investigate its discrete version using a lattice regularization. We also compare our model with the previous fermionic theories and study the structure of equal-time two-point functions in this setup.

\subsection{$\mathcal{L}_{\mathrm{I}}$ and $\mathcal{L}_{\mathrm{II}}$ Theories}
A 2$d$ free fermionic version of Lifshitz theory for generic integer values of $z$ has been introduced in \cite{Hartmann:2021vrt} and later on a similar family of theories for generic $d$ was introduced in \cite{Mozaffar:2021nex}. Hereafter we will denote these two family of theories with index I and II respectively. Theory II is well-defined for odd values of $z$ while for even $z$ careful consideration is needed. Our focus in this section and in most of this paper is on 2$d$ where the extension of theory II to even values of $z$ is straightforward. The Lagransian density for these theories are given by
\begin{align}
\mathcal{L}_{\mathrm{I}}&=\bar{\Psi} (i\gamma^0\partial_0+ \gamma^1  (i\partial_1)^z-m^z)\Psi,
\label{LI}
\\
\mathcal{L}_{\mathrm{II}}&=\bar{\Psi} (i\gamma^0\partial_0+ i\gamma^1T^{z-1} \partial_1-m^z)\Psi,
\label{LII}
\end{align}
where $T=\sqrt{-\partial_1\partial^1}$.

Both $\mathcal{L}_{\mathrm{I}}$ and $\mathcal{L}_{\mathrm{II}}$ lead to the desired dispersion relation for Lifshitz theories $\omega(k)=\sqrt{k^{2z}+m^{2z}}$. Furthermore, as it has been addressed in \cite{Hartmann:2021vrt}, $\mathcal{L}_{\mathrm{I}}$ leads to a strange property for the vacuum state of theories corresponding to even values of $z$, namely the vacuum state in this case turns out to be a product state in the position basis. This leads for instance to vanishing entanglement between spatial subregions in these theories. To the best of our knowledge there is no physical justification for such a feature and we believe that this odd behavior arises from the asymmetry between odd and even values of $z$ in \eqref{LI}, namely for even values of $z$ the imaginary unit in the spatial derivative vanishes. One important massage of this paper is a solution to this problem via a careful treatment of the square root of the square Laplace operator, as is introduced in \eqref{LII}. 

The $T$ operator introduced in \cite{Montani:2012ve} has the following properties 
	\begin{align}\label{TProp}
		T\partial_i&=\partial_i T,\\
		T^{2n}&=\left(\sqrt{-\partial_1^2}\right)^{2n}=\left(-\partial_1^2\right)^{n},\\
		T^{2n+1}&=\left(\sqrt{-\partial_1^2}\right)^{2n+1} =\left(\sqrt{-\partial_1^2}\right)^{2n}\left(\sqrt{-\partial_1^2}\right)=\left(-\partial_1^2\right)^{n} T.
	\end{align}
Moreover, this operator acts in momentum space as 
	\begin{align}
		T^{2n} \exp(ikx)&=k^{2n} \exp(ikx),\\
		T^n \exp(ikx)&=|k|^n\exp(ikx).
	\end{align}
In what follows, for our 2$d$ theory we will use the following choice for the gamma matrices
\begin{eqnarray}
		\gamma^0=\begin{pmatrix}
			0	&1  \\
			1&0 
		\end{pmatrix},\qquad	\gamma^1=\begin{pmatrix}
			0&-1  \\
			1&0 
		\end{pmatrix},\qquad	\gamma^2\equiv\gamma^0\gamma^1=\begin{pmatrix}
			1&0  \\
			0&-1 
		\end{pmatrix}.
	\end{eqnarray}
For future convenience, the Hamiltonian density corresponding to $\mathcal{L}_{\mathrm{II}}$ is found to be
	\begin{align}
		\mathcal{H}_{\mathrm{II}}&=\pi_{\Psi} \dot{\Psi}-\mathcal{L}  =-\bar{\Psi} (i\gamma^1T^{z-1} \partial_1-m^z)\Psi.
	\end{align}
Note that we will suppress the index II in the following.

\subsection{Lattice Regularization}

In order to find the discretized version of \eqref{LII} we need to express higher derivative terms on a lattice. To do so, we choose the following definition for the first derivative, which preserves the hermiticity of the lattice Hamiltonian
	\begin{equation}\label{eq:dis}
		\partial_1\Psi_n=\frac{\Psi_{n+1}-\Psi_{n-1}}{2}.
	\end{equation}
Due to the presence of the $T$ operator for even values of $z$, the definition of this operator on the lattice is challenging. To overcome this challenge, we use momentum space representation which can be obtained by using the Fourier transform as $\Psi_n =\frac{1}{L}\sum_{p=0}^{L} c_k e^{-i k n}$ with $k=\frac{2\pi }{L}p$ and $p\in \{0,\cdots, L-1\}$. Hence, we can rewrite the corresponding expression for the first derivative on the lattice as
\begin{align}
\begin{split}
\partial_1\Psi_n=\frac{\Psi_{n+1}-\Psi_{n-1}}{2}
&=\frac12 \left(c_k(\omega)e^{i((n+1)k+\omega k)}-c_k(\omega)e^{i((n-1)k+\omega k)}\right)
\\
&=\frac12 \left(e^{ik}-e^{-ik}\right)\Psi_n=i\sin(k)\Psi_n\;.
\end{split}
\end{align}
Similarly for the second derivative we have
\begin{align}
\begin{split}
\partial_1(\partial_1\Psi_n)=\frac{\partial_1 \Psi_{n+1}}{2}-\frac{\partial \Psi_{n-1}}{2}
&=\frac{\Psi_{n+2}-\Psi_{n}}{4} - \frac{\Psi_n-\Psi_{n-2}}{4}
\\
&=\left(\frac{1}{4} \left(e^{2 i k}+e^{-2 i k}\right)-\frac{1}{2}\right)\Psi_n=-\sin^2(k) \Psi_n.
\end{split}
\end{align}
In general, the $z$-th derivative yields
	\begin{equation}
		(\partial_1)^z\Psi_n=i^{-z} (-\sin(k))^z\Psi_n.
	\end{equation}
Using the above relations, we can simply find the action of $T$ operator in momentum space as follows
	\begin{eqnarray*}
		&&T^2 \Psi_n=-\partial_1^2\Psi_n=\sin^2(k) \Psi_n,\nonumber\\
		&&T \Psi_n=|\sin(k)| \Psi_n.
	\end{eqnarray*} 
	Now it is straightforward to find the corresponding expression for the Hamiltonian on the lattice
	\begin{equation}
		H_{\mathrm{II}}=\sum_{n}\Psi_n^{\dagger}\left( -\gamma^{2}f_{\mathrm{II}}(k)+m^z\gamma^0\right)\Psi_n,
	\end{equation}
	where

\begin{eqnarray}\label{fnewmodel}
f_{\mathrm{II, odd}}(k)=-(\sin(k))^z,\qquad\qquad f_{\mathrm{II, even}}(k)=-|\sin(k)|(-\sin(k))^{z-1}.
\end{eqnarray}
The hermiticity of the Hamiltonian is guaranteed by ${\gamma^0}^{\dagger}=\gamma^0$, ${\gamma^2}^{\dagger}=\gamma^2$, and $f_{\mathrm{II}}(k)$ being a real function. Next, we diagonalize this model by treating the two components of the fermionic field separately as 
$\psi_{k}=\left(u_k,d_k\right)^{T}.$ In this representation, the Hamiltonian takes the following form	
	\begin{align}
		H_{\mathrm{II}}=&\int dk  \left(   
		(d_k^{\dagger}d_k-u_k^{\dagger}u_k)
		f_{\mathrm{II}}(k)+ m^z 
		(u_k^{\dagger}d_k+d_k^{\dagger}u_k)
		\right).
	\end{align}
	We consider the following Bogoliubov transformations
	\begin{equation}\label{Bogoliubov}
		u_k=\cos \frac{\theta_k}{2}b_k+i \sin\frac{\theta_k}{2} b^{\dagger}_{-k},\qquad\qquad   d_k=\sin\frac{\theta_k}{2} b_k -i\cos\frac{\theta_k}{2}b_{-k}^{\dagger},
	\end{equation}
	with
	\begin{equation}
		\tan \theta_k=\frac{m^z}{-f_{\mathrm{II}}(k)},\qquad\qquad \cos \theta_k=-\frac{f_{\mathrm{II}}(k)}{\sqrt{m^{2z}+f_{\mathrm{II}}(k)^2}},
	\end{equation}
	where
	$\{ b_k^{\dagger},b_{k'} \}=\delta_{kk'}$. Finally, the diagonal Hamiltonian can find as below	
	\begin{equation}\label{eq:HIIdis}
		H_{\mathrm{II}}=\int dk \sqrt{m^{2z}+f_{\mathrm{II}}(k)^{2}} \left(   
		b_k^{\dagger}	b_k+b_{-k}^{\dagger}b_{-k}
		\right).
	\end{equation}
It is worth to note that this discretization introduced in Eq. \eqref{eq:dis} is different from what has been introduced in reference \cite{Mozaffar:2021nex}. In that reference forward difference operator has been used for all derivative orders together with the standard procedure to end up with a Hermitian Hamiltonian. The diagonalization follows from the same aforementioned steps by replacing $f_{\mathrm{II}}$ by $f^*_{\mathrm{II}}$ where   
\begin{equation}
f^*_{\mathrm{II}}(k)=\left(2\sin\frac{k}{2}\right)^z \cos\frac{z k}{2}.\label{f2}
\end{equation}
Let us also mention that for the case of \eqref{LI}, the authors of \cite{Hartmann:2021vrt} have used the discretization Eq. \eqref{eq:dis}. The Hamiltonian has the same form as \eqref{eq:HIIdis} with $f_{\mathrm{II}}(k)$ replaced by
\begin{equation}
f_{\mathrm{I}}(k)= (-\sin(k))^z\,.
\end{equation}
Note that  $f_{\mathrm{I}}$, $f_{\mathrm{II}}$, and $f^*_{\mathrm{II}}$ all lead to the desired Lifshitz dispersion relation, $\omega(k)=\sqrt{k^{2z}+m^{2z}}$, in the continuum limit.

Since we are interested in the two point correlators to compute entanglement measures, hereafter the expressions with $f(k)$ are valid for all three $f_{\mathrm{I}}(k)$ corresponding to $\mathcal{L}_{\mathrm{I}}$, and $f_{\mathrm{II}}(k)$, $f^*_{\mathrm{II}}(k)$ corresponding to different discretizations of $\mathcal{L}_{\mathrm{II}}$.
  
\subsection*{Two-point Function}
It is relatively straightforward to find the fermionic field's two-point function as follows
	\begin{eqnarray}\label{2pointfunction}
		C_{rs}=\langle   \Psi_s \Psi_r^{\dagger}\rangle=\frac{\delta_{rs}}{2} \mathbb{1}+\frac{1}{4\pi}\int_{-\pi}^{\pi} \frac{d k}{\omega_{k}} e^{ik(r-s)} \begin{pmatrix}
			-f(k)  &  m^z\\
			m^z& f(k)
		\end{pmatrix} .
	\end{eqnarray}
In the following we will mainly investigate the behavior of entanglement measures for $f_{\mathrm{II}}(k)$, while we present and compare the results with $f_{\mathrm{I}}(k)$ and $f^*_{\mathrm{II}}(k)$ to highlight the corresponding physical features.

Moreover, we also examine the behavior of logarithmic negativity for finite temperature state where the corresponding two-point function is given as follows	
\begin{eqnarray}\label{2pointfunctionthemal}
C_{rs}=\frac{\delta_{rs}}{2} \mathbb{1}+\frac{1}{4\pi}\int_{-\pi}^{\pi} \frac{d k}{\omega_{k}} e^{ik(r-s)} \begin{pmatrix}
			-f(k)  &  m^z\\
			m^z& f(k)
		\end{pmatrix} \tanh{\frac{\beta \omega_k}{2}}.
	\end{eqnarray}
As we will see studying the above expression in the low and high temperature limits will be illuminating. Indeed, a simple 
calculation gives the two-point function in the high temperature limit for odd values of $z$ as follows
\begin{eqnarray}\label{eq:2pntTodd}
	C^{\mathrm{odd}}_{rs}=\frac{\delta_{rs}}{2} \mathbb{1}+\frac{\beta }{2}\frac{1}{4\pi}
	\frac{2 i^z z! \sin (\pi  (r-s))}{2^z (r-s-z) \left(\frac{1}{2} (r-s-z+2)\right)_z}
	 \begin{pmatrix} 
		-1  &  0\\
		0& 1
	\end{pmatrix}+\cdots
\end{eqnarray}
and for the even values of $z$
\begin{eqnarray}\label{eq:2pntTeven}
	C^{\mathrm{even}}_{rs}=\frac{\delta_{rs}}{2} \mathbb{1}+\frac{\beta }{2}\frac{1}{4\pi}
	\frac{2 i^{z+1} z! (\cos (\pi  (r-s))-1)}{2^z (r-s-z) \left(\frac{1}{2} (r-s-z+2)\right)_z}
	 \begin{pmatrix} 
		-1  &  0\\
		0& 1
	\end{pmatrix}+\cdots
\end{eqnarray}
where $(\bullet)_{x}$ stands for the Pochhammer symbol.

We are now basically equipped with what we need to study the entanglement measures which will come in section \ref{sec:Numerics}. Before getting into our numerical results in the next subsections we first further analyse the difference between $\mathcal{L}_{\mathrm{I}}$ and $\mathcal{L}_{\mathrm{II}}$ and proceed by presenting a symmetry based argument which will be used to justify some of our results for Lifshitz fermions.       

\subsection{$\mathcal{L}_{\mathrm{I}}$ versus $\mathcal{L}_{\mathrm{II}}$ at the Fixed Point}
\label{sec:fIvsfII}
Before examining the full $z$ dependence of entanglement measures, we would like to study their specific behaviors in the massless regime where the theory respects scaling symmetry. As we will see this study plays an important role in our analysis in what follows. Indeed $f_{\mathrm{I}}(k)$, $f_{\mathrm{II}}(k)$, and $f^*_{\mathrm{II}}(k)$ have distinct behaviors in the massless regime which reflect the main features of each theory and the corresponding regularization.

\begin{figure}[h]
\centering
{\includegraphics[width=0.46\linewidth]{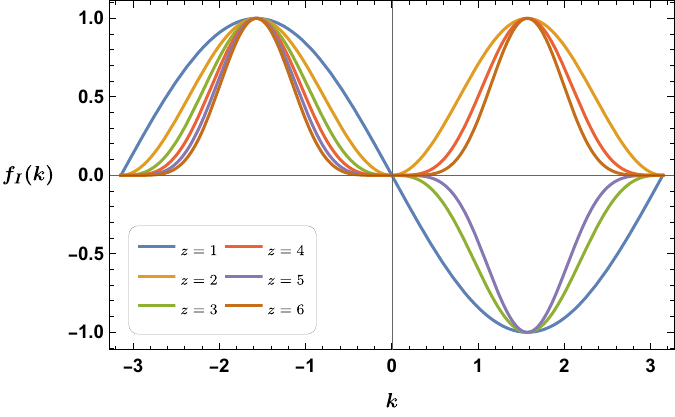}}\quad	{\includegraphics[width=0.46\linewidth]{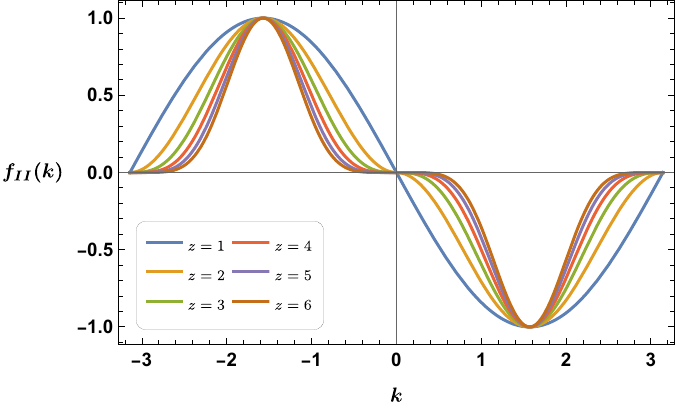}}
\caption{The behavior of $f_{\mathrm{I}}(k)$ (left) and $f_{\mathrm{II}}(k)$ (right) for several values of the dynamical exponent. For odd $z$ both models have the same behavior and so is the two-point functions. $f_{\mathrm{I}}(k)$ is always positive for even values of $z$ leading to an empty Fermi surface. Moreover the zeros of $f_{\mathrm{II}}(k)$ are the same for all values of $z$, and hence the two-point function in the massless case is $z$-independent.}
\label{fk}
\end{figure}

Eq. \eqref{2pointfunction} for $m=0$ simplifies to
\begin{eqnarray}\label{crsmassless}
C_{rs}=\frac{\delta_{rs}}{2} \mathbb{1}+\frac{1}{4\pi}\int_{-\pi}^{\pi} dk \;e^{ik(r-s)} \begin{pmatrix}
			-1  &  0\\
			0& 1
		\end{pmatrix} \frac{f(k)}{|f(k)|}.
\end{eqnarray}
We can see that, not the exact value of $f(k)$, but its sign between the zeros of it matters for this correlation function. The authors of \cite{Hartmann:2021vrt} have shown that for the case of $f_{\mathrm{I}}(k)$, for odd values of $z$ this sign changes at $k=n\pi$, thus in this case a straightforward calculation gives
	\begin{eqnarray}\label{oddzC}
		C_{rs}=\frac{1}{2}  \begin{pmatrix}
			\delta_{rs}-\frac{i \left(e^{i \pi  (r-s)}-1\right)}{2 \pi(r-s)}&  0\\
			0&\delta_{rs}+\frac{i \left(e^{i \pi  (r-s)}-1\right)}{2 \pi(r-s)}
		\end{pmatrix}\,. 
	\end{eqnarray}
On the other hand, for even values of $z$ the corresponding two-point function simplifies as follows
	\begin{eqnarray}\label{crsmasslesszeven}
		C_{rs}=\begin{pmatrix}
			0  &  0\\
			0&1
		\end{pmatrix} \delta_{rs}\,.
	\end{eqnarray}
Interestingly, the eigenvalues of the above matrix are either 0 or 1 and hence using Eq. \eqref{EE}, the entanglement entropy literally vanishes for even values of the dynamical exponent.
Moreover, for $f_{\mathrm{II}}(k)$, using Eq. \eqref{2pointfunction} we see that the result for the odd \textit{and even} values of $z$ are the same as Eq. \eqref{oddzC}.
In figure \ref{fk}, we compare the behavior of $f(k)$ for $\mathcal{L}_{\mathrm{I}}$ and $\mathcal{L}_{\mathrm{II}}$ where it manifestly shows that the Fermi surface ($f(k)=0$) is empty for even values of $z$ for $\mathcal{L}_{\mathrm{I}}$ but quite non-trivial for $\mathcal{L}_{\mathrm{II}}$.\footnote{We thank Masaki Oshikawa for bringing this point to our attention.} This structure manifestly shows why $\mathcal{L}_{\mathrm{I}}$ leads to a product state for the vacuum state of even values of $z$, while there is no counterpart for $+k$ modes inside the Fermi surface to be entangled with. It is worth noting that although this result correspond to the lattice regularized theories, exactly the same structure holds in the continuum limit of $\mathcal{L}_{\mathrm{I}}$ and $\mathcal{L}_{\mathrm{II}}$.

An important comment is about the complicated structure of $f^*_{\mathrm{II}}(k)$. Unlike $f_{\mathrm{I}}(k)$ and $f_{\mathrm{II}}(k)$ which have a single zero at $k=0$, $f^*_{\mathrm{II}}(k)$ accommodates $2(z-1)$ extra zeros. As a result the this regularization is distinct from $f_{\mathrm{I}}(k)$ and $f_{\mathrm{II}}$ by being explicitly $z$-dependent and not being bounded by the $z=1$ result. In this case the Fermi surface is also non-trivial and the vacuum state is an entangled state. We will address the implications of this point in section \ref{sec:Numerics}.

Our results show that for integer values of $z$ the two-point functions are all the same as the relativistic case. This structure is very different from the similar structure for Lifshitz invariant scalar theories \cite{MohammadiMozaffar:2017nri}. As we show in the next section, this structure also interestingly affects the entanglement structure. In the next section we try to further investigate the physics behind this specific behavior of the fermionic two-point function from a symmetry viewpoint.
	
\subsection{Two-point Function: Symmetry Argument}\label{symmlif}
It is a well-known fact that at a field theory fixed-point the typical scaling is given by Lifshitz scaling, \textit{i.e.}, Eq. \eqref{eq:LifScaling}. The goal of this subsection is to show that using this scaling symmetry with a reasonable assumption fixes the form of the two-point function of a fermionic field in Lifshitz theory. 

Various features of nonrelativistic symmetries have been studied in the literature, \textit{e.g.}, see \cite{Henkel:1993sg,Henkel:2003pu,Stoimenov:2005dk,Hartnoll:2009sz,Figueroa-OFarrill:2022kcd}. At a Lifshitz fixed point the symmetry algebra of the field theory will contain the generators of rotations $M_{ij} $, spatial translations $P_i$, time translation $H$ and dilatation $D$. In particular in 2$d$ we have
\begin{align}
		H=\partial_t, \qquad
		P=\partial_x, \qquad
		D=-zt\partial_t-x\partial_x\,.
\end{align}
The corresponding nontrivial commutation relations are
	\begin{eqnarray}
	[D,P]=P,\hspace*{2cm}[D,H]=zH.
		\end{eqnarray}
Moreover, under infinitesimal coordinate transformations, the fields transform as follows
\begin{align}
\begin{split}
		[H,\phi(x,t)]&=\partial_t \phi(x,t),\\
		[P,\phi(x,t)]&=\partial_x \phi(x,t),\\
		[D,\phi(x,t)]&=zt \frac{\partial \phi(x,t)}{\partial t}+\Delta  \phi (x,t)+ x \frac{\partial\phi(x,t)}{\partial x},
\end{split}
\end{align}
where we have used $\phi'(x',t')=\lambda^{-\Delta}\phi(x,t)$. Our goal is to apply these transformations to the (massless) fermionic fields. Indeed in this case the corresponding two-point function splits for the independent components of the fermionic field that each can be treated as a scalar field. Now we define
	\begin{eqnarray}
		F=F(x_a,x_b;t_a,t_b)=\langle\phi_a(x_a,t_a)\phi_b^*(x_b,t_b)\rangle,
	\end{eqnarray}
where $\phi_a(x_a,t_a)$ denotes a chiral component of the fermion field. Translation invariance implies that $F = F(x, \tau )$ where $x = x_a-x_b$ and $\tau = t_a-t_b$. 	Further, scale invariance which is generated by $D$ requires that 
	\begin{align}
		\left(zt_a \partial_a + x_a D_a +\Delta_a +zt_b \partial_b + x_b D_b +\Delta_b\right) F(x;\tau)&=0.
	\end{align}
We would like to recast the above equation in a more simplified form. We do so by defining $\partial_a=-\partial_b=\partial_t$ and $D_a=-D_b=D_x$ which yields 
\begin{align}\label{2pteq}
\left(z\tau \partial_\tau + x D_x +\Delta_a+\Delta_b\right) F(x;\tau)=0.
\end{align}
The general form of the two-point function can be found by solving the above equation. Of course, from dimensional analysis we can deduce 
	\begin{eqnarray}
		F(x,\tau)=x^{-\left(\Delta_a+\Delta_b\right)} G\left(\frac{x^z}{\tau}\right),
	\end{eqnarray}
where $G(u)$ is an arbitrary function.  
In what follows for simplicity we consider the equal time two-point function and thus $t_b=t_a$. We also assume $\Delta_b=\Delta_a\equiv\Delta$.\footnote{Indeed, this choice is consistent with our expectations from the general results for the two-point functions in the relativistic limit corresponds to $z=1$.} Under these assumptions, eq. \eqref{2pteq} yields
	\begin{align}
		\left(x D_x +2\Delta\right) F(x)=0,
	\end{align}
and the corresponding solution becomes
	\begin{eqnarray}
		F(x)=\frac{C}{x^{2\Delta}},
	\end{eqnarray}
where $C$ is a constant which can be set to unity after a proper normalization. Hence, the equal time two-point function depends on the dynamical exponent through the scaling dimension. Indeed, for the bosonic field, where $\Delta=\frac{1-z}{2}$, the two-point function becomes
	\begin{eqnarray}
		F(x)= C\, x^{z-1},
	\end{eqnarray}
with a nontrivial dependence on the dynamical exponent. On the other hand, for the fermionic field, where $\Delta=\frac12$, the two-point function becomes
	\begin{eqnarray}\label{2ptferm}
		F(x)=\frac{C}{x},
	\end{eqnarray}
which is \textit{independent} of the dynamical exponent. This behavior corresponds to both theories defined in Eq. \eqref{LI} and \eqref{LII}. It is also possible to write the above two-point function as the form reported in the previous section. Although, this expression is not well defined for $x_b=x_a$, we can find a more general expression which holds even in this specific case. To do so, we note that the delta function is a solution to eq. \eqref{2pteq}. Indeed, a straightforward calculation gives\footnote{Here we use the properties of delta function to show that
	\begin{equation*}
		x\delta(x)=0\longrightarrow \partial_x(x\delta(x))=0=\delta(x)+x\partial_x\delta(x)\longrightarrow\delta(x)=-x\partial_x\delta(x).
	\end{equation*}
}
	\begin{align}
		\left(x D_x +2\Delta\right) \delta(x)= x D_x  \delta(x)+2\Delta \delta(x)=-  \delta(x)+2\Delta \delta(x)=0
	\end{align}
where in the last equality we used $\Delta_a=\Delta_b=\frac12$. Now the general solution of Eq. \eqref{2pteq} up to a coefficient would be a linear combination of the solution given in Eq. \eqref{2ptferm} and the delta function as
	\begin{eqnarray}
		F(x_a-x_b)=c\,\delta(x_a-x_b)+\frac{d}{x_a-x_b},
	\end{eqnarray}
where $c$ and $d$ are constants. The above expression shares the same form with the two-point function we found in Eq. \eqref{2pointfunction}.

	\section{Numerical Results}\label{sec:Numerics}
	
In this section, we numerically investigate the entanglement measures mainly in 2$d$ for the aforementioned models. We provide a detailed numerical analysis and examine how entanglement measures depend on the dynamical exponent for different states and subregion configurations. We compare these measures across various aspects which enables us to gain insight into distinct features of our models.  We address the possibility of introducing a c-function defined in terms of entanglement entropy.

Through this section we always consider translational invariance in our lattice models by either considering an infinite lattice or applying periodic boundary conditions on a finite lattice. We would like to note that an appropriate factor to avoid fermion doubling problem is considered all over this section. 
	
\subsection{Entanglement Entropy}
Plugging the eigenvalues of the corresponding covariance matrix into Eq. \eqref{EE} we numerically evaluate the entanglement entropy.

\subsubsection*{Vacuum State of $\mathcal{L}_{\mathrm{II}}$}
\begin{figure}[h]
		\centering
		{\includegraphics[width=0.46\linewidth]{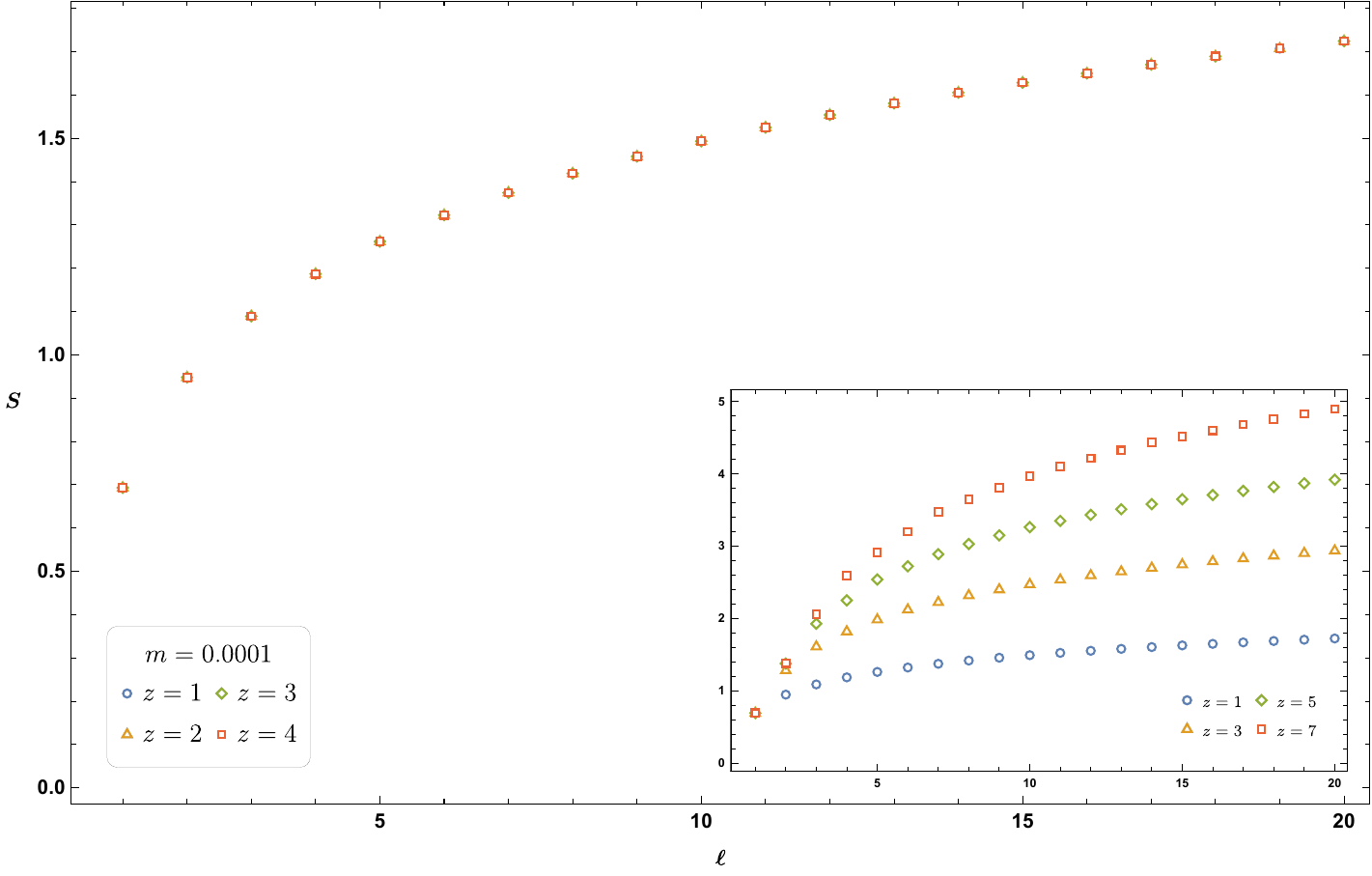}}\quad	{\includegraphics[width=0.46\linewidth]{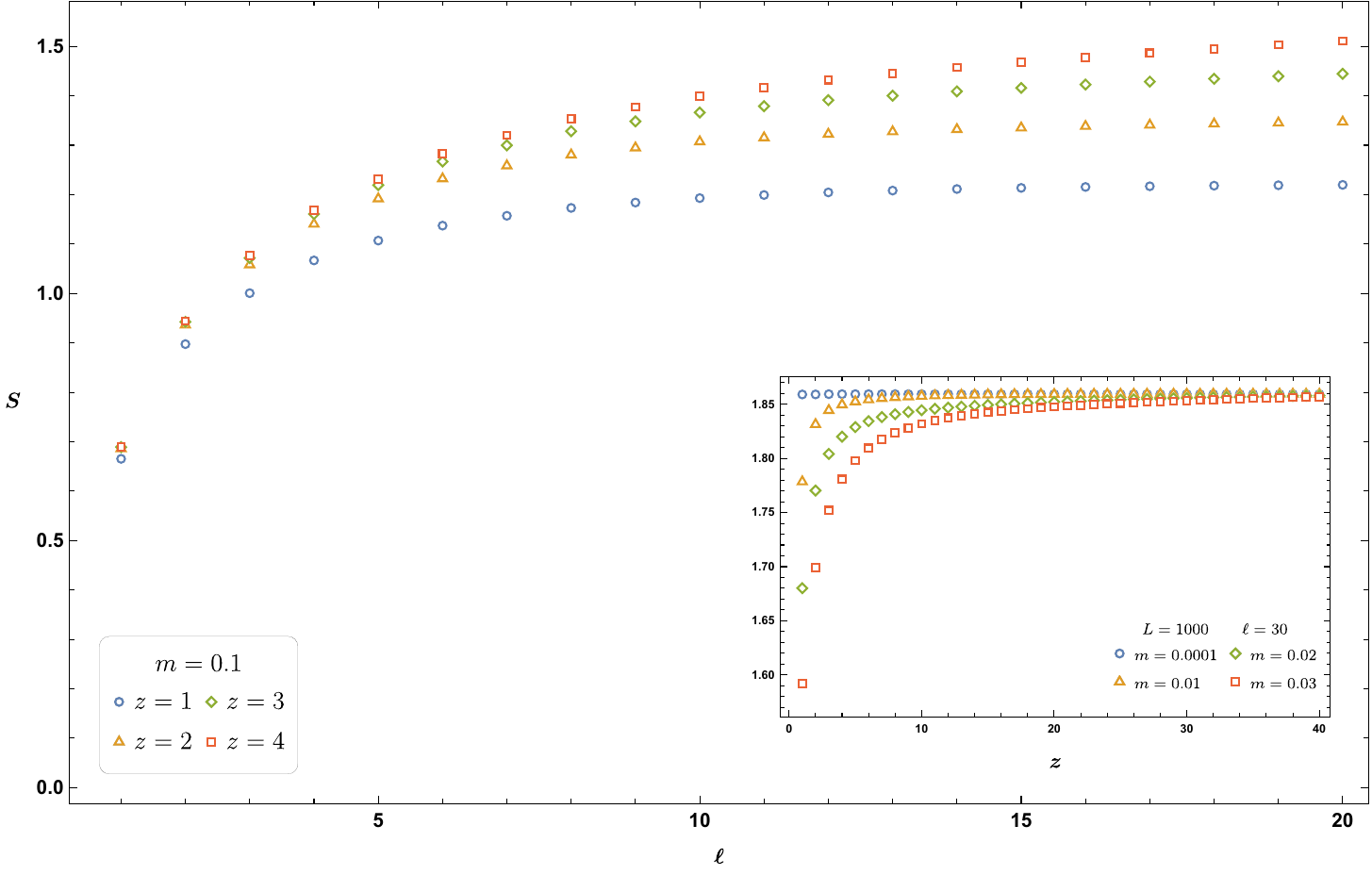}}
		\caption{Left: Entanglement entropy for a single interval in the massless regime as a function of the interval length $\ell$ for $f_{\mathrm{II}}(k)$. The inset shows the same plot corresponding to $f^*_{\mathrm{II}}(k)$ where there is no upper bound for EE. Right: Entanglement entropy in the massless regime as a function of $\ell$ for $f_{\mathrm{II}}(k)$. The inset shows how the massless result plays the role of an upper bound in massive theories for larger values of the dynamical exponent. All results correspond to infinite spatial direction.}
		\label{newmodelEE}
	\end{figure}

Figure \ref{newmodelEE} shows the entanglement entropy as a function of the length of the entangling region for model $f_{\mathrm{II}}(k)$ in $L\rightarrow \infty$ limit. The left panel shows that the entanglement entropy does not depend on the dynamical exponent in the vanishingly small mass limit. As a result there is \textit{no difference} between relativistic and nonrelativistic fermions in the massless (scale-invariant) regime.

In contrast, in Lifshitz scalar theories entanglement entropy monotonically increases with $z$ \cite{MohammadiMozaffar:2017nri, MohammadiMozaffar:2017chk}. The dynamical exponent can be interpreted as a measure for the correlation length, inspired from the interaction terms in the lattice model. This picture explains the observed behavior in the scalar theories. On the lattice this effect may lead to the emergence of volume law behavior for entanglement entropy when $z\gg n_A$.

With regard to figure \ref{newmodelEE}:
\begin{itemize}
\item
The behavior of massless $f_{\mathrm{II}}(k)$ fermions in the left panel is consistent with what we expect in the continuum limit from the structure of the two-point function (see section \ref{sec:fIvsfII}). Specifically, the $z$-independence of the entanglement entropy is inherited from the two-point function. 

\item
Massless $f_{\mathrm{II}}(k)$ fermions do \textit{not} follow the bosonic picture, that monotonically increases by $z$, but rather experience a strong upper bound preventing the amount of quantum correlations to exceed that of the $z=1$ case, namely $\frac{1}{3}\log n_A$. This upper bound is \textit{much} smaller than the intrinsic upper bound for fermions, namely $n_A\log 2$ for $n_A\gg1$.

\item
The inset of the left panel shows that the massless $f^*_{\mathrm{II}}(k)$ fermions do follow the bosonic picture and there is \textit{no} upper bound for the entanglement entropy. To get more insight about the difference between $f^*_{\mathrm{II}}(k)$ and $f_{\mathrm{II}}(k)$ one should note that in the latter $|f_{\mathrm{II}}(k)|\leq 1$ but it is easy to check that such a bound does not exist for $f^*_{\mathrm{II}}(k)$. The larger the value of the dynamical exponent, the larger $|f^*_{\mathrm{II}}(k)|$. Based on this key difference, regularization $f^*_{\mathrm{II}}(k)$ does \textit{not} reproduce the $z$-independent structure in the massless continuum limit. 

\item
The right panel shows the well-understood fact that the mass parameter decreases the entanglement entropy. Further, for the massive case increasing the dynamical exponent results in increase of this measure. This behavior is consistent with the massive scalar fields.

\item
The inset of right panel shows a trade off between the mass parameter and the dynamical exponent for $f_{\mathrm{II}}(k)$ fermions. The mass term decreases the entanglement entropy due to the decrease of the correlation length, while $z$ increases the entanglement entropy through increasing the correlation strength. The $z$-independent scale invariance result plays the role of an upper bound for this quantity in massive theories.
\end{itemize}


\subsubsection*{Vacuum State of $\mathcal{L}_{\mathrm{I}}$ versus $\mathcal{L}_{\mathrm{II}}$}
\begin{figure}[h]
\centering
{\includegraphics[width=0.46\linewidth]{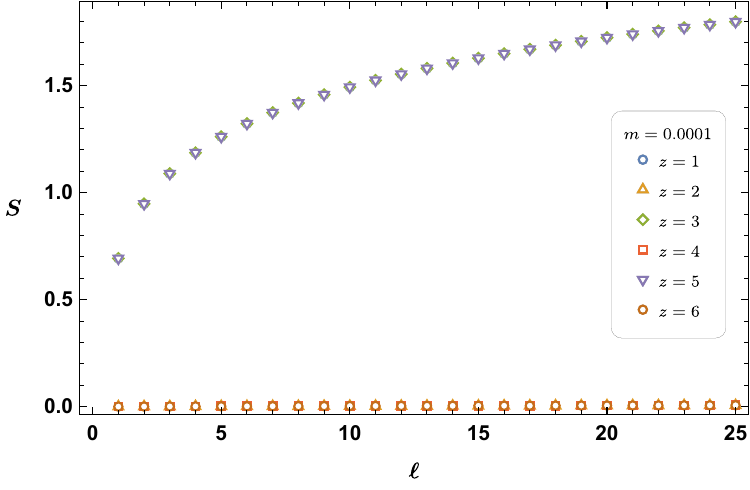}}	\quad	{\includegraphics[width=0.46\linewidth]	{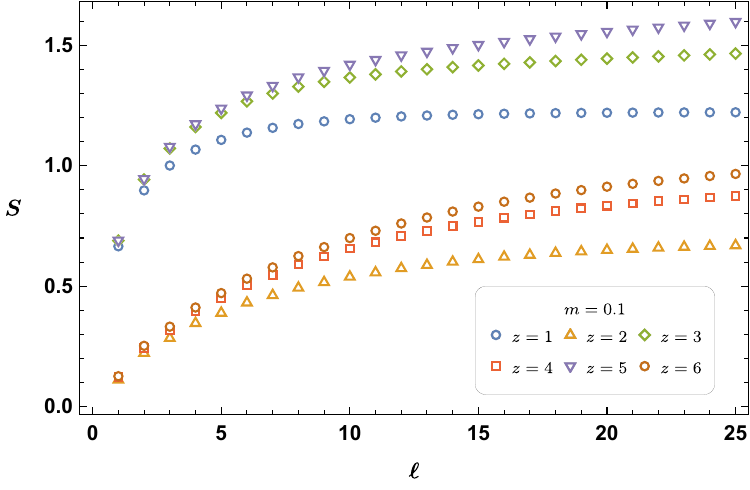}}
\caption{Entanglement entropy for $f_{\mathrm{I}}(k)$ in the massless (left) and massive (right) regimes for several values of the dynamical exponent.}
\label{vanmodel}
\end{figure}

Let us now turn to compare the entanglement entropy from $f_{\mathrm{II}}(k)$ with the previously known results from $f_{\mathrm{I}}(k)$. The left panel of figure \ref{vanmodel} corresponds to $m\to0$ limit. For odd values of the dynamical exponent, there is no surprise that entanglement entropy for $f_{\mathrm{I}}(k)$ coincides with that for $f_{\mathrm{II}}(k)$ illustrated in figure \ref{newmodelEE}. However, for even values of $z$, there is a peculiar property. As mentioned in \cite{Hartmann:2021vrt}, the entanglement entropy in the massless limit becomes zero due to the structure of the vacuum state which is a product state for spatial degrees of freedom (see Eq. \eqref{crsmasslesszeven}). The right panel corresponds to the massive case where entanglement entropy is non trivial for both even and odd values of the dynamical exponent. Moreover, for odd $z$ we show that the effect of the mass parameter is qualitatively similar to what we find in figure \ref{newmodelEE}, but for even values of $z$ the entanglement entropy counterintuitively \textit{increases} with the mass parameter. 

%
\subsubsection*{Thermal State}
\begin{figure}[h]
		\centering
		\includegraphics[width=0.46\linewidth]{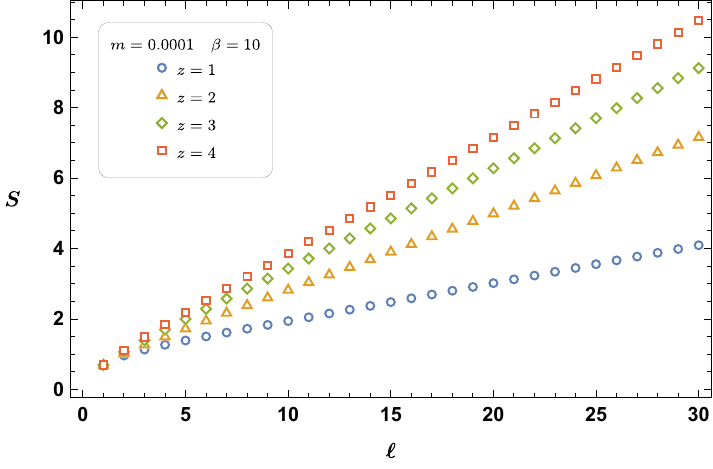}\quad\includegraphics[width=0.46\linewidth]{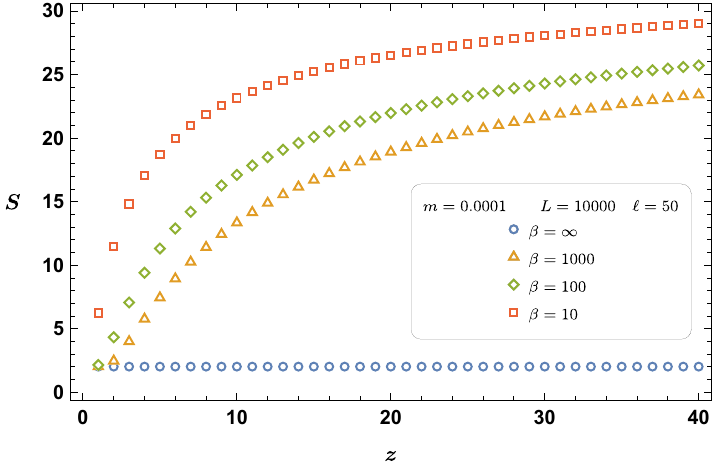}
		\	\caption{\textit{Left}: $S(\ell)$ for several values of the dynamical exponent at finite temperature. \textit{Right}: Entanglement entropy as a function of $z$ for different values of $\beta$. In both panels we consider a finite lattice with $L$ sites and the massless regime.}
		\label{EEtobeta}
	\end{figure}

Let us now examine the behavior of von Neumann entropy for a single interval in thermal state of $\mathcal{L}_{\mathrm{II}}$. Clearly this is not expected to be a quantum correlation measure. As we have already noted in this case we employ Eq. \eqref{2pointfunctionthemal} to find the corresponding two-point function for thermal states. The numerical results in the massless limit are summarized in figure \ref{EEtobeta}. Let us make a number of observations regarding to these plots. First, we note that in both plots, at finite temperature, entanglement entropy increases with the dynamical exponent. Next, at low temperature the dependence on $z$ becomes less pronounced such that in $\beta\rightarrow \infty$ limit, the entanglement entropy is independent of $z$ as expected. Moreover, the right panel is showing that the entropy of the subregion is saturating at some value $\sim n_A \log 2$, the natural upper bound on entropy. 
	
\subsubsection*{3$d$ versus 2$d$: $z$-dependence of the Entanglement Entropy}	
In this part we digress our main focus to look at the behavior of $\mathcal{L}_{\mathrm{II}}$ in 3$d$. The goal is to shed light on the unique properties of the fermionic Lifshitz theory in 2$d$. In two spatial dimensions, the two-point functions can be found as (see appendix \ref{App1})
	\begin{eqnarray}
	\langle \Psi^{\dagger}_{i,k} \Psi_{j,l}\rangle=\frac12 \delta_{ij}\delta_{kl}-\frac{1}{2L^2}\sum_{k_1,k_2} \frac{m \gamma^0+f(k_1)\gamma^0 \gamma^1 +f(k_2)\gamma^0 \gamma^2}{ \sqrt{m^2+f(k_1)^2+f(k_2)^2}}e^{i(k_1(i-j)+k_2 (k-l))}.  
\end{eqnarray}
In this case, we choose the following representation for the gamma matrices
\begin{equation}
		\gamma^0=	\begin{pmatrix}
			0&1  \\
			1& 0
		\end{pmatrix},\qquad
		\gamma^1=	\begin{pmatrix}
			0&1  \\
			-1& 0
		\end{pmatrix},
		\qquad
		\gamma^2=	\begin{pmatrix}
			i&0  \\
			0& -i
		\end{pmatrix}.
\end{equation}
With this choice the two-point function takes following form
\begin{eqnarray}\label{2ptfunc2d}
	\langle \Psi^{\dagger}_{i,k} \Psi_{j,l}\rangle=\frac12 \delta_{ij}\delta_{kl}-\frac{1}{2L^2}\sum_{k_1,k_2} \frac{e^{i(k_1(i-j)+k_2 (k-l))}}{ \omega_k}\begin{pmatrix}
		-f(k_1)&&  m^z-if(k_2) \\
		m^z +if(k_2)&& f(k_1) \\
	\end{pmatrix},  
\end{eqnarray}
where $\omega_k=\sqrt{m^{2z}+f(k_1)^2+f(k_2)^2}$ and we consider $f(k_i)=f_{\mathrm{II}}(k_i)$ in our numerical analysis.

\begin{figure}[h]
		\centering
		\includegraphics[width=0.5\linewidth]{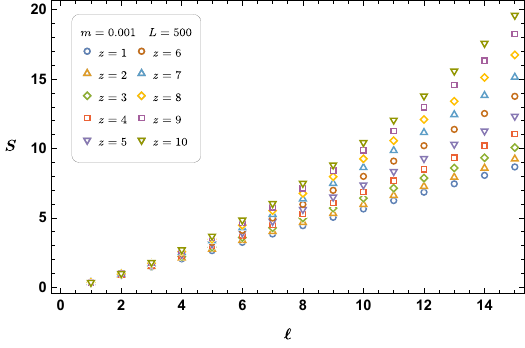}
		\caption{Entanglement entropy for a square entangling region with side length $\ell$ in the $m\to0$ regime.}
		\label{d2fermionnew}
	\end{figure}

By employing the generalization of the correlator method to higher dimensions we work out the entanglement entropy in this case were the results are shown in figure \ref{d2fermionnew} for a square entangling region in the $m\to0$ limit. Interestingly, we see that entanglement entropy monotonically \textit{increases} with the dynamical exponent. From Eq. \eqref{2ptfunc2d}, we see that in 3$d$, the corresponding two point function depends on $z$ even in the massless regime which is a strong evidence that $z$-independence is an intriguing feature related to the scaling dimension of the fermion field peculiar to 2$d$. From the data presented in figure \ref{d2fermionnew}, one can verify from the quadratic $\ell$-dependence of the entanglement entropy that for large values of $z$, namely when $z\gg n_A$ where $n_A$ is the number of sites on a side of the square region, entanglement entropy scales with the volume of the subregion. 

However, we should note that numerical results in a 3$d$ generalization of $\mathcal{L}_{\mathrm{I}}$ indicate that although entanglement entropy is not zero for even values of the dynamical exponent, but it is not a monotonic function of $z$. Indeed, we find $S(z_{\rm even})<S(z_{\rm odd})$ for $z_{\rm even}>z_{\rm odd}$ for certain regions in the parameter space. 

\subsection{Logarithmic Negativity}
In this section we mainly consider logarithmic negativity for mixed states as well as mutual information which were defined  in section \ref{sec:intro}. We calculate logarithmic negativity for fermionic Gaussian states using fermionic partial transpose by Eq. \eqref{LN} and also address the behavior of mutual information that can be calculated by plugging Eq. \eqref{EE} into Eq. \eqref{MI}. 

\subsubsection*{Vacuum State}
\begin{figure}[h]
		\centering
		{\includegraphics[width=0.46\linewidth]{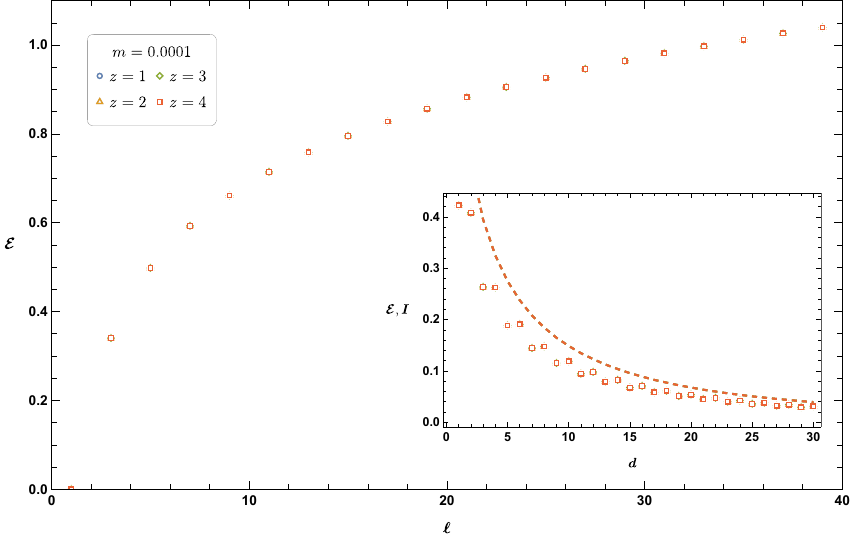}}	\quad	{\includegraphics[width=0.46\linewidth]{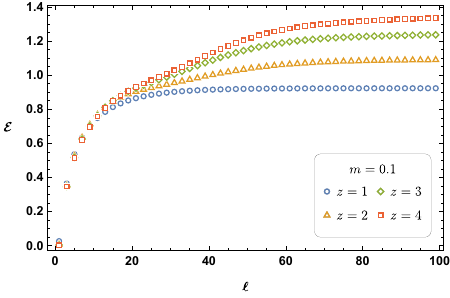}}
		\caption{Logarithmic negativity for two adjacent subregions with equal width $\ell$ for (left) massless and (right) massive regime. The inset of the left panel shows the results for both logarithmic negativity and mutual information for disjoint symmetric subregions. Here we consider $L\rightarrow\infty$ limit and the full state is the vacuum state.}
		\label{adjacentLN}
	\end{figure}

We consider two intervals with same width $\ell$ and separation $d$ on the vacuum state and trace out the complement. Figure \ref{adjacentLN} shows logarithmic negativity between these symmetric intervals. With regard to this figure:
\begin{itemize}
\item
The left panel shows that logarithmic negativity for \textit{adjacent} intervals does not depend on $z$ in the massless regime similar to the entanglement entropy for a connected interval in the vacuum state. Here we again recall that it was shown in \cite{MohammadiMozaffar:2017chk} that in a Lifshitz scalar theory $\mathcal{E}$ monotonically increases with the dynamical exponent which is again in contrast with the current results for the fermions.
\item
The inset of this panel corresponds to a configuration that these symmetric intervals are at a finite distance $d$. The inset also includes the results for the mutual information in the same configuration that is indicated by dashed lines. Both these measures are again independent of $z$ and decreases as one increases the separation between the intervals, expected in local theories.
\item
In the right panel, we have plotted the massive regime for adjacent intervals. In this case logarithmic negativity grows with the region width and saturates to a finite value in the large $\ell$ limit. Moreover, logarithmic negativity increases with $z$ but there is still an upper bound that saturates to.
\end{itemize}

\subsubsection*{Thermal State}
Let us now examine the behavior of logarithmic negativity at finite temperature states. In order to interpret our results, we would like to recall that negativities take their maximum value for pure states (\textit{e.g.}, logarithmic negativity for a pure state is equal to Renyi entropy with index $n=1/2$ which is even larger than the von Neumann entropy) and vanish for maximally mixed states. A quick extrapolation of this property leads us to expect negativities to be decreasing as the mixedness of the state increases. For instance it is well known that the temperature corrections to negativities always take a negative sign, see g.e. \cite{Calabrese:2014yza}.  

\begin{figure}[h]
		\centering
		\includegraphics[width=0.45\linewidth]{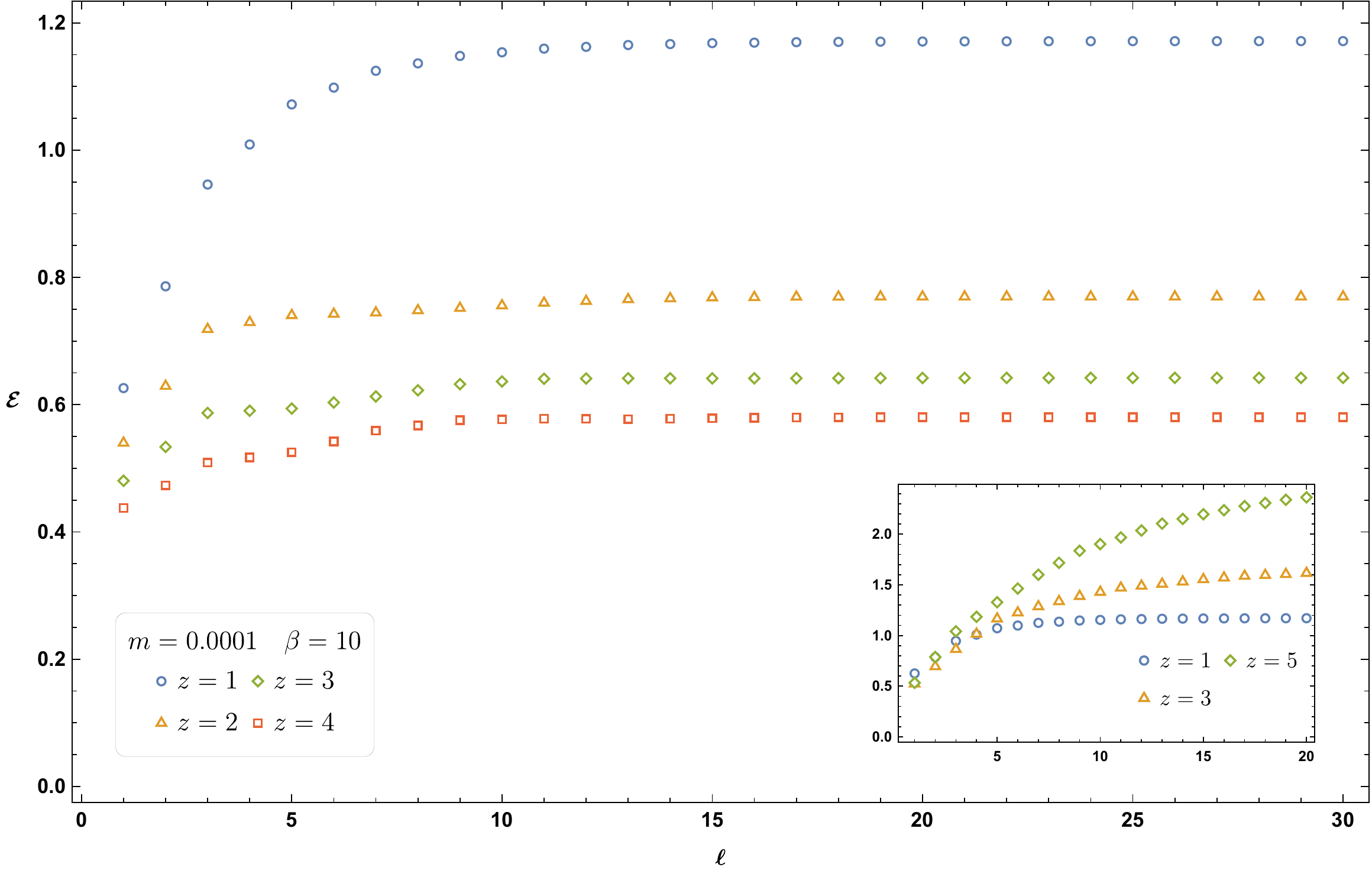}\quad\includegraphics[width=0.45\linewidth]{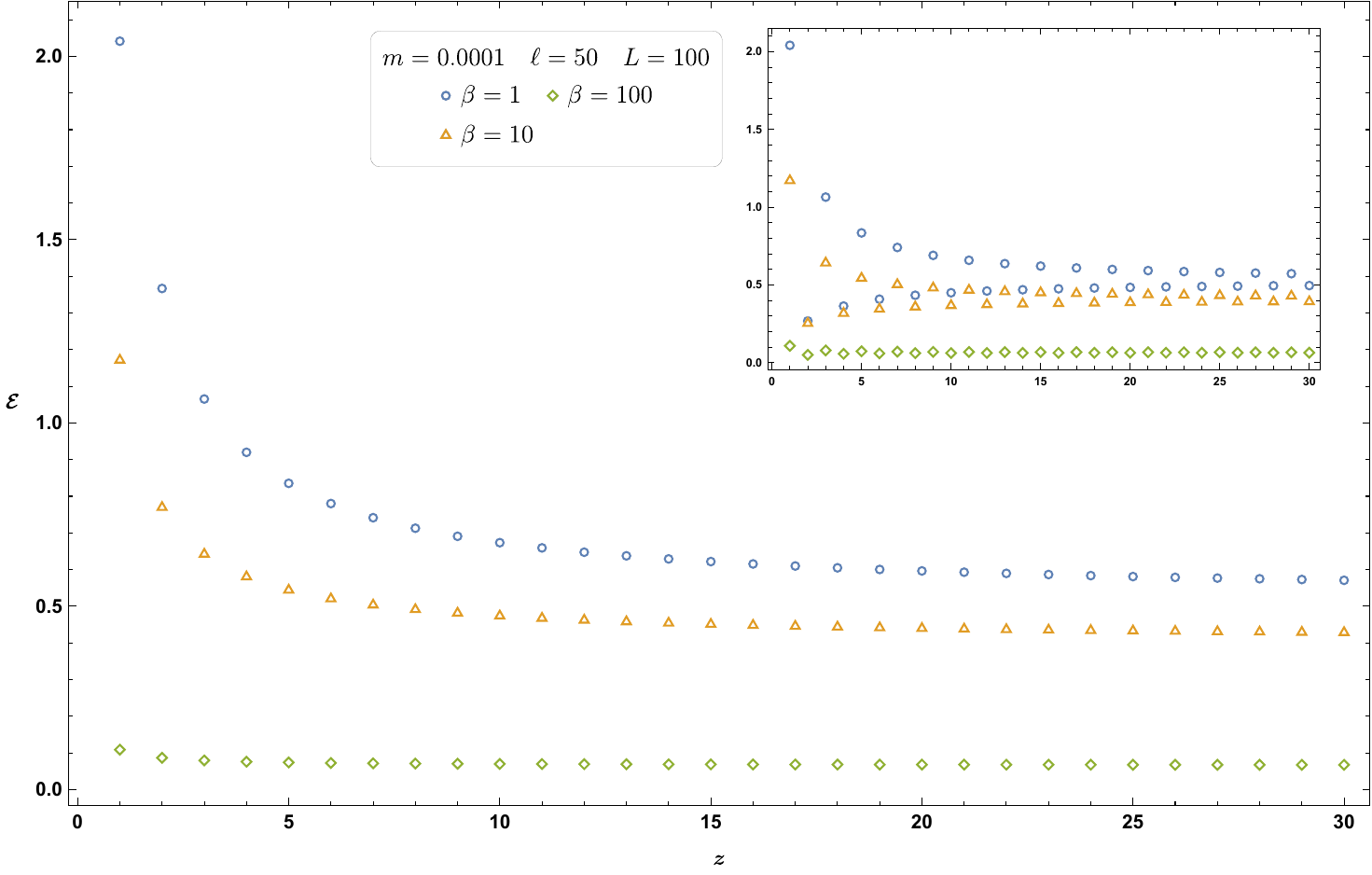}\quad
		\caption{\textit{Left}: Logarithmic negativity as a function of subregion width for several values of the dynamical exponent with  $\beta=10$. The inset of the left panel shows the results for $f^*_{\mathrm{II}}(k)$ . \textit{Right}: Logarithmic negativity as a function of the dynamical exponent for several values of the temperature.  The inset of the right panel shows the results for $f_{\mathrm{I}}(k)$ . :  In all panels we consider single intervals in small mass regime with $L=100$. }
		\label{LNtobeta}
\end{figure}

In figure \ref{LNtobeta} we illustrate our numerical results where we have considered a single interval in a finite temperature state and the massless regime. With regard to figure \ref{LNtobeta}:

\begin{itemize}
\item
From the left plot $\mathcal{E}(\ell)$ is a monotonically increasing function that almost saturates to some finite value. We believe this saturation happens due to the locality of the theory and the correlation scheme which is $z$-independent.   

\item
The right panel shows that as expected from physical grounds, logarithmic negativity decreases with the temperature for a fixed value of $z$. As the temperature increases, the number of relevant terms in the expansion of the thermal state increases. This increases the mixedness of the state though logarithmic negativity decreases.

\item
The right panel also shows that logarithmic negativity is decreasing as a function of $z$.\footnote{For different aspects about the effect of $z$ in quantum correlations in terms of odd entropy see \cite{Mollabashi:2020ifv}.} The $z$-dependence behavior is rather complicated compared to $\beta$-dependence. Although $\omega(k)$ is an increasing function of $z$, the eigenvalues of the mixed state also depend on the fermionic occupation number. Strictly speaking, focusing on mode $k$, increasing $z$ decreases the eigenvalue corresponding to the unoccupied state while increases the one corresponding to the occupied state. The net effect of all modes forms the the number of relevant terms in the thermal state expansion. This can be explained with the behavior of the two-point function. In this fermionic \textit{local} theory at finite temperature, as one can easily verify from Eqs. \eqref{eq:2pntTodd} and \eqref{eq:2pntTeven}, the net effect of increasing $z$ on the thermal correlation function is \textit{decreasing} the correlation. Though it is not surprising that quantum correlations captured by logarithmic negativity should also decrease.
\end{itemize}

	\subsection{Entropic $c$-function}
	
\begin{figure}[h]
\centering
\includegraphics[width=0.47\linewidth]{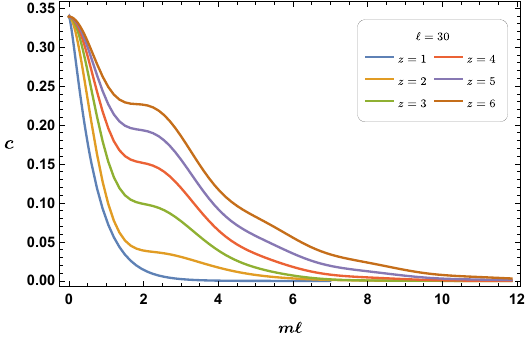}
\includegraphics[width=0.46\linewidth]{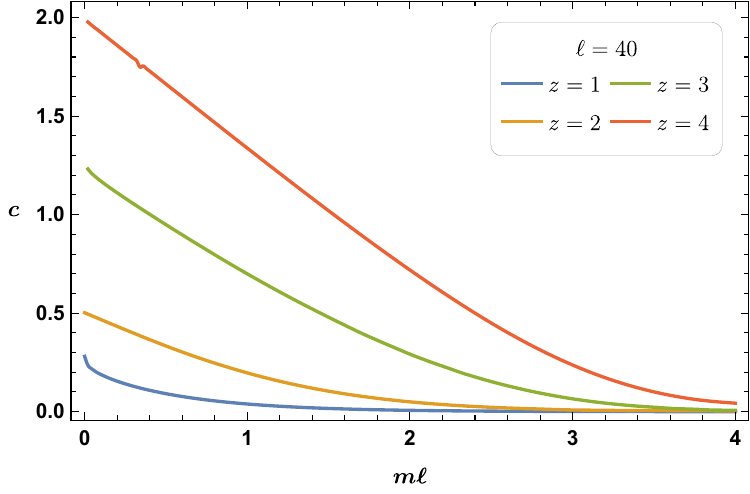}
\caption{$c$-function as a function of mass for different values of the dynamical exponent for fermionic (left) and scalar  (right) fields.}
		\label{cfun}
	\end{figure}

In this section we consider another interesting quantity which estimates the monotonic readjustment of entanglement under the RG flow, the so-called (entropic) $c$-function \cite{Casini:2006es,Casini:2009sr}. In a 2$d$ relativistic theory, this quantity is defined as follows 
	\begin{eqnarray}\label{cfunction}
		c(\ell)=\ell \frac{dS}{d\ell},
	\end{eqnarray}
which at the corresponding fixed points coincides with the coefficient of the universal logarithmic term in the expansion of the entanglement entropy for a subsystem of width $\ell$. Indeed, this function monotonically decreases along the RG flow from UV to IR which is a manifestation of $c$-theorem. Because the mass term is relevant we expect that $c$-function decreases as the mass increases. Here we assume that the same definition as Eq. \eqref{cfunction} holds\footnote{A justification for this assumption: the definition of \eqref{cfunction} is based on two assumptions, 1) the existence of a light-cone structure in the theory, 2) a logarithmic leading term for the entanglement entropy. The existence of the light-cone structure has been addressed in Lifshitz invariant free scalar theories in \cite{MohammadiMozaffar:2018vmk} for lattice models and in \cite{Mozaffar:2021nex} for the continuum limit. The same argument holds for free fermions as the argument is solely based on the form of the dispersion relation. With regard to the second assumption, for fermionic theories at the UV fixed point the entanglement entropy is $z$-independent and the relativistic result is valid for any $z$. For bosonic theories as long the entangling region (in the units of the UV cut off) is much larger than the dynamical exponent, the leading term of the entanglement entropy is logarithmic \cite{MohammadiMozaffar:2017nri, He:2017wla}.} for these non-relativistic theories and evaluate the corresponding $c$-function numerically. The results are collected in figure \ref{cfun} (left panel) where we show the $c$-function as a function of mass parameter. For any $z$, it starts at the same value of the relativistic case given by $c_{\rm rel}=\frac{1}{3}$, then decreases monotonically and saturates to zero in the large mass limit. A curious feature that we found is that the $c$-function has a plateau at some $m_\star$. Here we observe that as $z$ increases, the $c$-function approaches a constant value $\frac{1}{3}$. 

The right panel of figure \ref{cfun} illustrates the behavior of the $c$-function in the Lifshitz scalar theory for several values of the dynamical exponent. Our numerical results show that the number of effective number of degrees of freedom is increasing with $z$ while for a fixed $z$, as for the $z=1$ case, the theory is flowing to a massive theory in IR limit where the effective number of degrees of freedom gets arbitrarily small as the the IR scale is decreased.

In the Dirac fermion case for $z=1$, the behavior of the $c$-function coincides with that of the $z=1$ scalar theory. For larger values of the dynamical exponent the UV value of the $c$-function is $z$-independent, as the entropy was, and behavior through the flow is qualitatively similar to the scalar case. For arbitrary non-vanishing IR scale, the effective number of degrees of freedom gets is an increasing function of $z$ subject to the upper bound at the UV scale. However, as a natural result of the existence of such an upper bound, in the $z\rightarrow \infty$ limit the flow would be trivial since the introduced $c$-function would take a constant value $c=\frac{1}{3}$ all over the flow.



\section{Conclusions and discussions}

In this paper, we explored the entanglement structure in a new $(1+1)$-dimensional fermionic field theory whose UV fixed point respects Lifshitz scaling symmetry. We have carefully examined the behavior of several entanglement measures in static cases both for vacuum and mixed states including finite temperature states. We now proceed to summarize some interesting implications of our results and also some potential future directions that would be interesting to pursue.

\begin{figure}[h]
\centering
\includegraphics[width=0.46\linewidth]{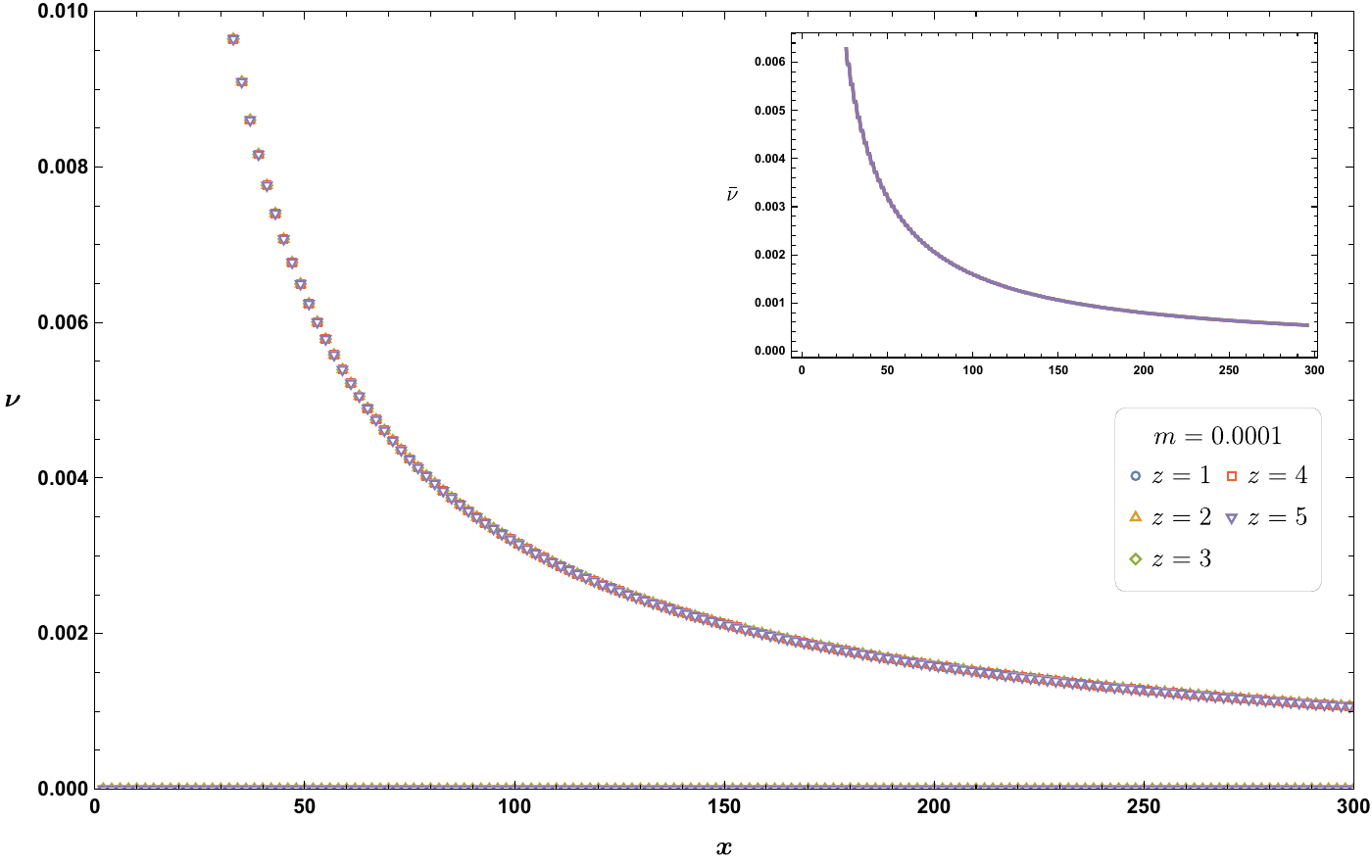}
\includegraphics[width=0.46\linewidth]{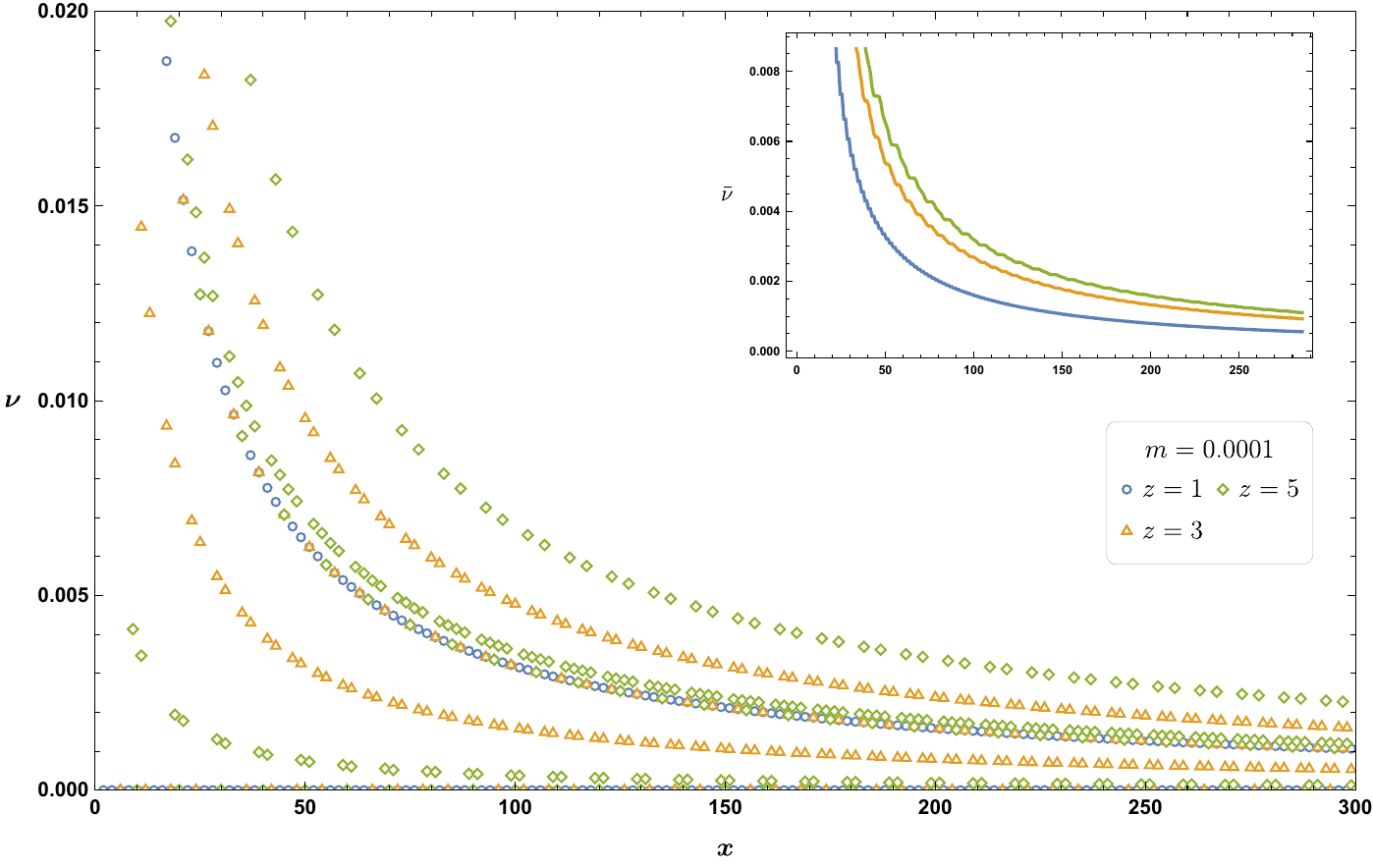}
\caption{Comparison of correlation strength in (left) $f_{\mathrm{II}}$ and (right) $f^*_{\mathrm{II}}$. The eigenvalue for the correlation matrix as a function of distance between two points $x$. There are a bunch of points with vanishing $\nu$ which correspond to some values with even value of $x$, suppressed in $f^*_{\mathrm{II}}$ as $z$ is increased. For a more clear comparison, in the insets we show the moving averaged value $\bar{\nu}$ for the same plot.}
\label{corfIIfIIs}
\end{figure}

Considering the entanglement entropy for a massless fermionic field (corresponds to the UV fixed point of the theory) in the vacuum state for $f_{\mathrm{II}}$, we have found that this measure is $z$-independent. Indeed, this behavior differs from the corresponding results for $f^*_{\mathrm{II}}$ which are previously reported in \cite{Mozaffar:2021nex}. We expect that this intriguing result can be explained in terms of the $z$-dependence of the fermionic field’s two-point function in Eq. \eqref{2pointfunction}. Figure \ref{corfIIfIIs} illustrates the eigenvalue of the correlation matrix as a function of distance between two lattice sites. As we can verify from this figure (left panel), the correlation function as a function of distance in $f_{\mathrm{II}}$ regularization follows the same pattern of the continuum limit in the sense that there is no $z$-dependence in the massless limit. On the other hand, from the right panel corresponding to $f^*_{\mathrm{II}}$, we see that there is a $z$-dependence behavior such that increasing the dynamical exponent, enhances the correlation. Further, this figure shows that there is no upper bound on the correlations in the $f^*_{\mathrm{II}}$ regularization.

%

Additionally, based on a symmetry argument and some physical assumptions we have shown that the two-point function for fermionic fields does not depend on the dynamical exponent and thus the $z$-independence of the entanglement entropy is inherited from this particular behavior. Further, our approach can be easily extended to higher dimensions where we have found that entanglement entropy monotonically increases with the dynamical exponent even in the massless regime. Thus we have shown that $z$-independence is an fascinating feature which particularly appears in $(1+1)$-dimensions.

Moreover, in order to gain further insights into certain properties of the entanglement structure of mixed states in fermionic theories, we also studied logarithmic negativity and an entropic $c$-function in a specific regime of the parameter space. Focusing on the $z$-dependence of logarithmic negativity between two subregions when the whole systems is in the vacuum state, we show that it does not depend on the dynamical exponent in the massless regime similar to the entanglement entropy for a single interval. Again, we can understand this intuitively in terms of the behavior of the fermionic two-point function. This behavior is in contrast with the previous results for the Lifshitz scalar theory where $\mathcal{E}$ monotonically increases with the dynamical exponent. On the other hand, in the massive case the logarithmic negativity increases with $z$ and then saturates from below to a finite value, depending on the mixedness of the state. 

Further, considering a single interval at finite temperature in the massless regime, $\mathcal{E}(\ell)$ is a monotonically increasing function that more or less saturates to a finite value in the large $\ell$ limit. Indeed, this specific behavior is due to the locality of the theory and the correlation scheme which is $z$-independent. Remarkably, our numerical results show that the logarithmic negativity is a monotonically decreasing function of the dynamical exponent. We have argued that this peculiar behavior is due to a competition between the $z$-dependence of $\omega(k)$ and the fermionic occupation number such that the net effect of increasing the dynamical exponent on the thermal correlation function is decreasing the total correlation. Hence, logarithmic negativity which captures quantum correlations also decreases. Moreover, regarding the $z$-dependence of the mutual information, our numerical analysis supply results qualitatively similar to observations made for the entanglement entropy as expected. 

We have also observed that employing the same relativistic definition yields a convenient (entropic) $c$-function which decreases along the RG flow. Specifically, our numerical results show that for arbitrary integer value of the dynamical exponent, the proposed function starts at the same value of the relativistic case and decreases monotonically through the flow before in the large mass limit. A curious feature that we have found is that the $c$-function has a plateau at some value of the mass parameter. 

The results of this paper are restricted to entanglement and correlation measures with Lifshitz fermions for static states. A straightforward extension of these results would be studying entanglement and correlation measures with Lifshitz fermions in out-of-equilibrium states, \textit{e.g.}, states followed by by quantum quenches. Another interesting question would be examining the behavior of other entanglement measures including entanglement of purification and odd entropy.

\section*{Acknowledgements}
We would like to thank Masaki Oshikawa, Saleh Rahimi-Keshari, Shinsei Ryu, Hassan Shapourian, Ken Shiozaki, and Erik Tonni for fruitful discussions. 
MV would like to thank the  School of Physics of IPM  for hospitality during this work. MV is also grateful to Mohsen Alishahiha for his supports.
MV is supported by Iran National Science Foundation (INSF) under project No.4023620.
AM would like to acknowledge support from ICTP through the Associates Programme (2023-2028) and for hospitality during the final stages of this work. 
Some numerical calculations of this work has been carried on facilities provided by the School of Physics, IPM. 
\begin{appendix}

		\section{Lifshitz fermionic model in $(2+1)$-dimensions}\label{App1}
		
The Lagrangian of the lifshitz fermionic model in $(2+1)$-dimensions is given as follows
		\begin{eqnarray}\label{lagrangian2+1}
			\mathcal{L}=\bar{\Psi} (i\gamma^0\partial_0+ i\gamma^xT^{z-1}_{(x)} \partial_x+ i\gamma^yT^{z-1}_{(y)} \partial_y-m)\Psi,
		\end{eqnarray}
where $T_{(x)}=\sqrt{-\partial_x^2}$ and $T_{(y)}=\sqrt{-\partial_y^2}$. It is straightforward to show that in this case the corresponding Hamiltonian becomes
		\begin{align}
			\mathcal{H}&=\pi_{\Psi} \dot{\Psi}-\mathcal{L}  =-\bar{\Psi} (i\gamma^xT^{z-1}_{(x)} \partial_x+i\gamma^yT^{z-1}_{(y)} \partial_y-m)\Psi.
		\end{align}
In $(1+2)$-dimensions, we choose the following majorana representation for the gamma matrices
	\begin{equation}
		\gamma^0=	\begin{pmatrix}
			0&1  \\
			1& 0
		\end{pmatrix},\qquad
		\gamma^1=	\begin{pmatrix}
			0&1  \\
			-1& 0
		\end{pmatrix},
		\qquad
		\gamma^2=	\begin{pmatrix}
			i&0  \\
			0& -i
		\end{pmatrix}.
	\end{equation}
A similar derivation to the one presented for $(1+1)$-dimensions gives the Hamiltonian density on the square lattice as follows
		\begin{eqnarray}
			\mathcal{H}=-\sum_{n_x=0,n_y=0}^{N}\bar{\Psi}_{\vec{n}} (iT^{z-1}(\gamma^1 \partial_1+\gamma^2 \partial_2)-m)\Psi_{\vec{n}}.
		\end{eqnarray}
Now using the Fourier transformations we can find the momentum space representation which yields
		
	\begin{eqnarray}
		\mathcal{H}	=\sum_{k_1,k_2=0}\psi^{\dagger}_k \begin{pmatrix}
			f(k_1)&m +i f(k_2) \\
			m-i f(k_2)&- f(k_1)
		\end{pmatrix}	\psi_k,
	\end{eqnarray}
	where $f(k)$ is the same as eq. \eqref{fnewmodel}. Further, we can diagonalize the above Hamiltonian by treating the two components of the fermionic field separately as 
	$
	\psi_{k}=
	\begin{pmatrix}
		u_k\\
		d_k
	\end{pmatrix}
	$. Using eq. \eqref{Bogoliubov}, the diagonalization constraint gives
	\begin{eqnarray}
		i f(k_1) \sin\theta_k -f(k_2)-i m \cos\theta_k=0.
	\end{eqnarray}
Moreover, solving the above equation	we can find $\sin\theta_k$ and $\cos\theta_k$ as follows 
		\begin{eqnarray}\label{sincos}
			\cos \theta_k&=&\frac{\sqrt{f(k_2)^2 \left(f(k_1)^2+f(k_2)^2+m^2\right)}+i m f(k_1)}{f(k_1)^2+f(k_2)^2},\nonumber\\
			\sin \theta_k&=&\frac{-\sqrt{f(k_1)^2 \left(f(k_1)^2+f(k_2)^2+m^2\right)}+i m f(k_2)}{f(k_1)^2+f(k_2)^2}.
		\end{eqnarray}
Finally, the diagonal Hamiltonian becomes
		\begin{equation}
			H=\sum_{k_1,k_2} \sqrt{m^{2z}+f(k_1)^{2z}+f(k_2)^{2z} } \left(   
			b_{\vec{k}}^{\dagger}	b_{\vec{k}}+b_{-{k}}^{\dagger}b_{-\vec{k}}
			\right)
		\end{equation}
In the next step, we can find the two-point function using the following equation
		\begin{equation*}
			\langle   \Psi_s^{\dagger} \Psi_r\rangle=\frac{1}{L^2}\sum_{k,k'} e^{ikr}e^{-ik's}\langle  \psi_{k'}^{\dagger}\psi_{k} \rangle.
		\end{equation*}
Employing the definition of Bogoliubov transformation, we can write the two point function in terms of $\sin\theta_k$ and $\cos\theta_k$ as follows
	\begin{eqnarray}
		\langle   \Psi_s^{\dagger} \Psi_r\rangle 
		&=& \frac{\delta_{rs}}{2} \mathbb{1}-\frac{1}{2L^2}\sum_{\vec{k}} e^{i\vec{k}(\vec{r}-\vec{s})} \begin{pmatrix}
			\cos \theta_{k}   &  \sin\theta_k\\
			\sin\theta_{k}& -\cos \theta_{k}
		\end{pmatrix}.
	\end{eqnarray}
Now, we can use eq. \eqref{sincos} to find the two-point function which yields		
	\begin{eqnarray}
		\langle \Psi^{\dagger}_{i,k} \Psi_{j,l}\rangle=\frac12 \delta_{ij}\delta_{kl}-\frac{1}{2L^2}\sum_{k_1,k_2} \frac{m \gamma^0+f(k_1)\gamma^0 \gamma^1 +f(k_2)\gamma^0 \gamma^2}{ \sqrt{m^2+f(k_1)^2+f(k_2)^2}}e^{i(k_1(i-j)+k_2 (k-l))}.
	\end{eqnarray}

\end{appendix}





\end{document}